\documentstyle[prb,multicol,aps,epsfig]{revtex}

\newcommand{\bq}{\begin{equation}}
\newcommand{\eq}{\end{equation}}
\newcommand{\bqa}{\begin{eqnarray}}
\newcommand{\eqa}{\end{eqnarray}}
\newcommand{\nn}{\nonumber \\}

\renewcommand{\narrowtext}{\begin{multicols}{2} \global\columnwidth20.5pc}
\renewcommand{\widetext}{\end{multicols} \global\columnwidth42.5pc}

\def\be		{\begin{equation}}
\def\ee		{\end{equation}}
\def\bea        {\begin{eqnarray}}
\def\eea        {\end{eqnarray}}
\def\bnn	{\begin{eqnarray*}}
\def\enn	{\end{eqnarray*}}

\begin{document}

\draft

\title{Role of spinon and spinon singlet pair excitations on phase transitions in $d-wave$ superconductors}
\author{Ki-Seok Kim, Jae-Hyeon Eom, Young-Il Seo, and Sung-Ho Suck Salk$^{1}$}
\address{Department of Physics,
Pohang University of Science and
Technology Pohang 790-784, Korea \nn 
${}^{1}$Korea Institute of Advanced Studies, Seoul 130-012, Korea}

\maketitle

\begin{abstract}
We examine the roles of massless Dirac spinon and spin singlet pair excitations 
on the phase transition in $d-wave$ superconductors.
Although the massless spinon excitations in the presence of the spin singlet pair excitations 
do not alter the nature of the phase transition at $T = 0$, 
that is, the XY universality class,
they are seen to induce an additional attractive interaction potential between vortices,
further stabilizing vortex-antivortex pairs at low temperature for lightly doped high $T_c$ samples.
\end{abstract} 


\narrowtext

Recent thermal hall conductivity measurements\cite{ONG} suggest the existence of vortices in the pseudogap (PG) phase.
This implies that preformed pairs are present in the PG phase\cite{Preformed}.
In the PG phase, vortex-antivortex pairs remain broken to cause a state of globally incoherent but locally coherent 
Cooper pairs. Vortex induced phase transitions in underdoped region have been
an issue of great interest\cite{Nayak,Tesanovich,Ye,Herbut,Kleinert,Herbut_vortex}.
In this paper, by applying a duality transformation we extend the U(1) gauge Lagrangian\cite{D.H.Lee} obtained from the
slave-boson mean field theory\cite{Kotliar,Patrick} in order to examine
how vortex induced phase transitions in $d-wave$ superconductors at low energy
are affected by the presence of both the massless spinon and spinon singlet pair excitations.
In other study\cite{Kleinert,Herbut_vortex}, the flavor number of the massless Dirac fermions
without the spinon singlet pair excitations is shown to alter the nature of the phase transition.
According to this study\cite{Kleinert}, as the flavor number increases, 
the type $II$ superconductivity is preferred showing the second order phase transition
which deviates from the XY universality class.
In the case of small flavor number leading to the type $I$ superconductivity,
it becomes the first order transition owing to the strong fluctuations of the massless gauge field.
Our present study differs from other previous studies\cite{Kleinert,Herbut_vortex}
in that in our case the U(1) effective gauge field
of interest becomes massive as a result of the spinon singlet pair excitations.
Particle-hole excitations of the massless Dirac spinons
lead to the renormalized kinetic energy of the U(1) gauge field\cite{Tesanovich,Don.Kim,Franz} (the second term in Eq. [3]).
However, we find that 
the XY universality class is not altered despite the presence of the massless Dirac fermions (spinons). 
Thus it is irrespective of the flavor number as long as the effective gauge particle (Eq. [5] and Eq. [6])
remains sufficiently massive.
The U(1) Berry gauge field can emerge from coupling between the massless Dirac fermions and
the vortices\cite{Tesanovich,Ye,Herbut,Herbut_vortex}, which may affect
the XY universality class.
However, the Berry gauge field becomes massive owing to the charge fluctuations\cite{Ye} 
and as a consequence the effective dual Lagrangian
stays robust to maintain the XY universality class.
It is shown from the present study that the interaction potentials between vortices are
modified to bring an additional attractive interaction (Eq. [7] and Eq. [11]) as a consequence of the massive gauge field.

Our primary focus is to examine how low energy excitations affect
the vortex induced phase transition.
Here low energy excitations refer to the phase fluctuations
of both the spinon singlet pair and the single holon order parameter and the
massless Dirac spinon excitations near the $d-wave$ nodes of the spinon singlet pair.
Gauge field fluctuations are introduced to allow the presence of internal flux responsible for energy lowering.
Thus considering proper phase fluctuations (involved with
$\phi_{sp} = e^{i\theta_{sp}}$ and $\phi_b = e^{i\theta_{b}}$) for the spinon pairing order 
field and the single holon order field respectively, we rewrite
the low energy effective Lagrangian of Lee\cite{D.H.Lee} in compact form,
\bqa
&&{\cal{L}} = \frac{K_{b,\mu}}{2}|\partial_\mu\theta_b + a_\mu - A_\mu|^2 + \frac{K_{sp,\mu}}{2}|\partial_\mu\theta_{sp} + 2a_\mu|^2
\nn && + \psi_{1}^\dagger[\partial_\tau + v_F\tau^3i\partial_x + v_\Delta\tau^1{i}\partial_y]\psi_1 +
(1\rightarrow{2}, x\rightarrow{y}) \nn && + 
iJ_{f\mu}(\partial_\mu\theta_{sp} + 2a_\mu) + i\bar{\rho}_{sp}(\partial_\tau\theta_{sp}-2\partial_\tau\theta_b+2A_0),
\eqa
where $K_{b,\mu} \equiv (1/u_b, K_b, K_b)$ with $1/u_b$ ($\sim 1/t$), the compressibility and $K_b$ ($\sim 2t\chi_0\delta$),
the phase stiffness of the single holon field and $K_{sp,\mu} \equiv (1/u_{sp}, K_{sp}, K_{sp})$ with $1/u_{sp}$ 
($\sim 1/J$), the compressibility and $K_{sp}$ ($\sim J\Delta_0^2$), the phase stiffness of the spinon pair order field.
$\psi_{n\sigma} = \left (\begin{array}{c}
e^{-i\theta_{sp}/2}f_{n\sigma} \\ e^{i\theta_{sp}/2}\epsilon_{\sigma\sigma{'}}f_{n\sigma{'}}^{\dagger} \end{array} \right)$ is the
renormalized Nambu spinor associated with the $d-wave$ gap nodes $n$. 
$v_F$ ($\sim J\chi_0$) and $v_\Delta$ ($\sim J\Delta_0$) are
the fermi and gap velocities of the Dirac spinons respectively.
$J_{f\mu} = \frac{1}{2}(\sum_{n}\psi_{n\sigma}^\dagger\tau^3\psi_{n\sigma}, iv_F\psi_{1\sigma}^\dagger\psi_{1\sigma},
 iv_F\psi_{2\sigma}^\dagger\psi_{2\sigma})$ is the
three current of the spinon quasiparticles and $\bar{\rho}_{sp}$, the average density of spinon pairs.

By introducing the unitary gauge transformation $\tilde{a}_\mu = 2a_\mu + \partial_\mu\theta_{sp}$, we rewrite Eq. [1]
\bqa
&&{\cal{L}} = \frac{{K}_{b,\mu}}{8}|\partial_\mu\theta_p - \tilde{a}_\mu + 2A_\mu|^2 + \frac{K_{sp,\mu}}{2}\tilde{a}_\mu^2
+ iJ_{f\mu}\tilde{a}_\mu \nn
&& + \psi_{1}^\dagger[\partial_\tau + v_F\tau^3i\partial_x + v_\Delta\tau^1{i}\partial_y]\psi_1 +
(1\rightarrow{2}, x\rightarrow{y}) \nn && + i\bar{\rho}_{sp}(\partial_\tau\theta_{p} + 2A_0),
\eqa
where $\theta_p = \theta_{sp} - 2\theta_{b}$.
$\theta_p$ is the phase of the Cooper pair order parameter $\Delta_{Cooper}(k) = |\Delta^0_{Cooper}(k)|\phi_p$
where $\Delta_{Cooper}(k) = <c_{k\uparrow}c_{-k\downarrow}> = <b^*f_{k\uparrow}b^*f_{-k\downarrow}>
= <f_{k\uparrow}f_{-k\downarrow}><b^*>^2 = |\Delta_0(k)||<b^*>|^2\phi_{sp}{\phi_b^*}^2$ and thus $\phi_p = e^{i\theta_p} =
\phi_{sp}{\phi_b^*}^2 = e^{i\theta_{sp}}e^{-i2\theta_b}$.
$\tilde{a}_\mu$ is the massive effective gauge field associated with the phase fluctuations of spinon singlet pair order parameter and 
the original internal $U(1)$ gauge field $a_{\mu}$.
The mass of the effective gauge field $\tilde{a}_{\mu}$ is defined by the phase stiffness $K_{sp}$ of the spinon pairing order parameter
associated with the PG phase of the doped Mott insulator.
In the above equation, the fluctuating fields of $\theta_p$, 
$\tilde{a}_\mu$ and $\psi_n$, are U(1) gauge invariant, thus satisfying the Elitzur's theorem\cite{Elitzur}.

Integrating over the Nambu spinor fields and expanding the resulting logarithmic term upto
second order in $\tilde{a}_\mu$\cite{Tesanovich,Don.Kim,Franz},
we get an effective U(1) Lagrangian involved with 
the massive gauge field $\tilde{a}_\mu$,
\bqa
&&Z = \int D{\theta_p}D{\tilde{a}_\mu}e^{-{\int_0}^\beta d\tau \int d x^2 {\cal{L}} }, \nn
&&{\cal{L}} = \frac{\tilde{K}_{b,\mu}}{2}\Bigl|\partial_{\mu}\theta_p - \tilde{a}_\mu + 2A_\mu\Bigr|^2 +
\frac{N}{16}(\partial\times\tilde{a})\frac{1}{\sqrt{-\partial^2}}(\partial\times\tilde{a})
\nn && + \frac{1}{2}m_{a,\mu}^2\tilde{a}_\mu^2 + i\bar{\rho}_{sp}(\partial_\tau\theta_p + 2A_0),
\eqa
where $\tilde{K}_{b,\mu} = K_{b,\mu}/4$ and $m_{a,\mu}^2 = K_{sp,\mu}$.
N is the number of flavors (i.e., the number of nodal points) of the Dirac fermions.
The kinetic energy term (the second term) of the effective gauge field $\tilde{a}_\mu$ arises as a result of the 
massless excitations of the spinon quasiparticles (Dirac fermions)\cite{Tesanovich,Don.Kim,Franz}.
The Berry phase contribution $i\bar{\rho}_{sp}\partial_\tau\theta_p$ 
 is related to the Cooper pair boson density\cite{Fisher,NaLee}.

If we ignore the spinon quasiparticles and thus consider only the $N=0$ limit (which corresponds to 
the isotropic $s-wave$ superconductivity), it is obvious that the kinetic energy term of the gauge field
disappears. Integrating over the effective gauge field $\tilde{a}_{\mu}$ in Eq. [3], we obtain 
\bqa
{\cal{L}} = \frac{K_{p,\mu}}{2}|\partial_\mu\theta_p + 2A_\mu|^2 + i\bar{\rho}_{sp}(\partial_\tau\theta_{p} + 2A_0),
\eqa
 where ${K_{p,\mu}} = \frac{{\tilde{K}_{b,\mu}}{K_{sp,\mu}}}{{\tilde{K}_{b,\mu}} + {K_{sp,\mu}}}$.
This expression shows that the phase stiffness ${K_{p,\mu}}$ of the Cooper pair field is
in a reduced "mass" form and the usual logarithmic type of interaction between vortices arises.

The duality transformation\cite{Nayak,Ye,Fisher,NaLee,Na} of Eq. [3] with the introduction 
of vortex mass and self interaction terms leads to 
an effective Lagrangian for the vortex field,
\bqa
&&Z = \int {D\psi_{pV}}D{c_\mu}D{\tilde{a}_\mu}e^{-\int d x^3 {\cal{L}} }, \nn
&&{\cal{L}} = |(\partial + ic)\psi_{pV}|^{2} + m_{pV}^2|\psi_{pV}|^2 + \frac{u_{pV}}{2}|\psi_{pV}|^4 \nn && +
\frac{1}{2\tilde{K}_{b,\mu}}\Bigl|{\bf{\partial\times{c}}}\Bigr|^2 
+ i({\bf{\partial\times{c}}})_\mu\Bigl(\tilde{a}_\mu - 2A_\mu\Bigr) - \mu(\partial\times{c})_{\tau} \nn
&& + \frac{N}{16}(\partial\times\tilde{a})\frac{1}{\sqrt{-\partial^2}}(\partial\times\tilde{a})
 + \frac{1}{2}m_{a,\mu}^2\tilde{a}_\mu^2,
\eqa
where $m_{pV}^2 \sim \tilde{K}_{b} - \tilde{K}_{bc} \sim \delta - \delta_{c}$ with $\delta$, 
the hole doping concentration and $\delta_{c}$, the critical hole doping concentration
and $\mu = \frac{\bar{\rho}_{sp}}{\tilde{K}_{b,0}} = \tilde{u}_{b}\bar{\rho}_{sp}$.
$\psi_{pV}$ represents the Cooper pair vortex field
and ${c_\mu}$, the dual gauge field to mediate interactions between vortices. 
Noting that $\mu$ is analogous to an applied "magnetic field" $H_z$ and $(\partial\times{c})_{\tau}$ to $B_z$\cite{Fisher,NaLee},
the sixth term in Eq. [5], $-\mu(\partial\times{c})_\tau$\cite{Fisher,NaLee} which results
from the Berry phase term $i\bar{\rho}_{sp}\partial_\tau\theta_p$ is analogous 
to interaction energy $-H_{z}B_{z}$ associated with the vortex field.
We note that in case of $\delta < \delta_{c}$ vortex condensation occurs.

Integrating over the effective gauge field $\tilde{a}_{\mu}$, we get
\bqa
&&Z = Z_0^a\int{D\psi_{pV}}{D}{c_\mu}e^{-\int{d^3x} {\cal{L}}_{eff}} \nn
&&{\cal{L}}_{eff} = |(\partial + ic)\psi_{pV}|^{2} + m_{pV}^2|\psi_{pV}|^2 + 
\frac{u_{pV}}{2}|\psi_{pV}|^4 \nn && +
\frac{1}{2\tilde{K}_{b,\mu}}\Bigl|{\bf{\partial\times{c}}}\Bigr|^2 
+ \frac{1}{2}(\partial\times{c})\frac{1}{\frac{N}{8}\sqrt{-\partial^2} + m_{a,\mu}^2}(\partial\times{c}) \nn
&& - i\partial\times{c}\cdot{2A} - \mu(\partial\times{c})_{\tau}, \\
&&Z_0^a = \int{D}{\tilde{a}_\mu}e^{-\int{d^3x}
\frac{N}{16}(\partial\times\tilde{a})\frac{1}{\sqrt{-\partial^2}}(\partial\times\tilde{a})
+ \frac{1}{2}m_{a,\mu}^2\tilde{a}_\mu^2}.  \nonumber
\eqa
The fifth term represents an additional kinetic energy of the dual
 gauge field resulting from coupling of the massless Dirac spinon field 
to the Cooper pair field via the massive effective gauge field $\tilde{a}_{\mu}$.
Not considering the external magnetic field $\mu$, despite the contribution of the massless Dirac fermions (as shown in the fifth term)
the nature of the XY universality class will not be affected as long as 
the mass of the gauge field, that is, the phase stiffness $K_{sp}$ of the spinon singlet pair order parameter
is substantially large; the lower temperature, the larger $K_{sp}$ in the PG phase.

By considering a static case of the vortex field 
we calculate the dual gauge field propagator to obtain
the interaction potential between vortices.
The dual field propagator involved with the fourth and fifth terms in Eq. [6] is obtained,
\bqa
&&P_{ij}(q) = P(q)\Bigl(\delta_{ij} - \frac{q_i{q_j}}{q^2}\Bigr) \nn
&&P(q) = \frac{\tilde{K}_{b}(q + \frac{8}{N}m_a^2)}{q^2(q+\frac{8}{N}m_a^2+\frac{8}{N}\tilde{K}_{b})} \nn
&& = K_p\frac{1}{q^2} + \frac{K_b}{4}z_J\frac{1}{q(q+M^2)},
\eqa
where $K_{p} = \frac{K_{b}K_{sp}}{K_{b} + 4K_{sp}}$, $M = \sqrt{\frac{8}{N}(K_{sp} + K_b/4)}$ and 
$z_J = \frac{{K}_b}{{K}_b + 4K_{sp}}$. $i, j = 1, 2$ denotes the space component index $(x,y)$.
In the real space, the above equations lead to the total interaction potential 
between a vortex and an antivortex,
\begin{eqnarray}
&&V(x-x') = 2{\pi} K_pln|x-x'|\nn && - \pi^2\tilde{K}_bz_J(StruveH_0(M^2|x-x'|) - Y_0(M^2|x-x'|)) \nn
&& \sim  2\pi{K_p}ln|x-x'| + \pi^2\tilde{K}_bz_Jln|x-x'| , \\ && \mbox{ for } {M^2|x-x'|<<1} \nn
&&\sim 2{\pi}K_pln|x-x'| - \frac{2\pi\tilde{K}_bz_J}{M^2}\frac{1}{{|x-x'|}} ,\\ && \mbox{ for } {M^2|x-x'|>>1} \nonumber
\end{eqnarray}
where $Y_0(x)$ is the zeroth order Bessel function of the second kind and
 $StruveH_0(x)$ is the zeroth order Struve function\cite{Math}.
The additional interaction between vortices
shows a power law decay at large separation distances between vortices while it is logarithmic in nature at short separations.
In Fig. 1, we plot this interaction potential as a function of distance in the underdoped region\cite{Plots}.
The correction term yields a large modification 
to the logarithmic interaction energy particularly at short
distances, showing additional attractive interactions between vortices. 
It becomes more effective
at lower temperature and causes stronger vortex-antivortex binding.
However, at sufficiently large distances and thus at high temperature it becomes 
negligibly small.
The vortex-antivortex unbinding transition temperature 
$T_{KT}(\delta)$ as a function of hole doping $\delta$ is expected to linearly scale with 
the phase stiffness $K_p$ of the Cooper pair field. $K_p$ is linearly dependent on $\delta$ particularly in the lightly doped region.
On the other hand, the strength of the additional interaction shows a nonlinear (quadratic) dependence of $\delta$
owing to the linear doping dependence of $K_{b} (\sim \delta)$ and $z_{J} (\sim \delta)$ in the lightly doped region\cite{Plots}.

To grasp the origin of the additional attractive interaction between vortices in a different angle,
we now take a different procedure.
Integrating over the effective gauge field $\tilde{a}_{\mu}$ first in Eq. [2],
we obtain the effective Lagrangian\cite{D.H.Lee} of the U(1) gauge invariant particles
containing the Doppler energy shift term\cite{Doppler},
 $z_{\mu}J_{f\mu}\partial_{\mu}\theta_p$,
\bqa
&&Z = \int D{\psi}D{\theta_p} e^{-{\int_0}^\beta d\tau \int d x^2 {\cal{L}} }, \nn
&&{\cal{L}} = \frac{K_{p,\mu}}{2}|\partial_\mu\theta_p - 2A_\mu|^2 -iz_{\mu}J_{f\mu}(\partial_\mu\theta_p - 2A_\mu) \nn && +
\psi_{1}^\dagger[\partial_\tau + v_F\tau^3i\partial_x + v_\Delta\tau^1{i}\partial_y]\psi_1 +
(1\rightarrow{2}, x\rightarrow{y}) \nn && + \frac{1}{2}\frac{J_{f\mu}^2}{\tilde{K}_{b,\mu} + K_{sp,\mu}},
\eqa
with $K_{p,\mu} = \frac{K_{b,\mu}K_{sp,\mu}}{K_{b,\mu}+4K_{sp,\mu}} \equiv (1/u_p, K_p, K_p)$,
 the phase stiffness of the Cooper pair order parameters and 
$z_{\mu} = \frac{K_{b,\mu}}{K_{b,\mu}+4K_{sp,\mu}} \equiv (z_\rho, z_J, z_J)$, 
the effective charge of the spinon quasiparticles\cite{D.H.Lee}.
The above Lagrangian is the low
energy Lagrangian of the $d-wave$ BCS theory with the doping dependent phase stiffness,
$K_p$ ($\sim \delta$) and the effective charge, $z_J$ ($\sim \delta$)\cite{D.H.Lee}.

 To see the effects of the Dirac fermions on the phase fluctuations of the Cooper pair fields, 
we integrate over the Dirac fermion fields ignoring the local interactions 
of the Dirac fermions and the temporal fluctuations in Eq. [10] to find
\bqa
&&{\cal{L}} = \frac{K_{p}}{2}|\nabla\theta_p - 2A|^2 + \frac{N}{16}z_{J}^2(\nabla\times\nabla\theta_p - 2\nabla\times{A})
\nn &&\times\frac{1}{\sqrt{-\nabla^2}}(\nabla\times\nabla\theta_p - 2\nabla\times{A}).
\eqa
The additional attractive interaction between vortices leads to the second term
in Eq. [9]\cite{Additional_interaction} as a result of
the supercurrent affected by the massless Dirac fermions 
owing to the Doppler shift term $z_{\mu}J_{f\mu}\partial_{\mu}\theta_p$. 
For a brief guidance we list differences in interaction potentials between vortices in the $s-wave$ and
the $d-wave$ superconductors in Table 1 and a comparison of the $d-wave$
BCS theory and the present theory in Table 2.
We note that in the limit of $q<<M^2$ in Eq. [7]
the interaction energy between vortices obtained by the $d-wave$ BCS formalism is the same as that obtained by
our theory. This limit corresponds to the
case in which the local interactions of the massless Dirac fermions are ignored in Eq. [10].
The additional interactions may affect the vortex lattice structure\cite{vortex_lattice}.
We believe that because our
vortex interaction terms reveal the hole doping dependence, it will be of great interest in
the future to examine how the vortex dynamics and lattice structure vary with hole
concentration.

We noted that 
the phase transition in underdoped cuprates falls into the XY universality class owing to the presence of 
the massive gauge field $\tilde{a}_{\mu}$ which results from the spinon singlet pair excitations. 
In addition, we showed that the massless 
Dirac fermions coupled to the phase fluctuations of the spinon singlet pairs (leading to the Cooper pairs)
result in the additional attractive interaction of a logarithmic behavior, $\frac{\pi}{2}\tilde{K}_bz_Jln|x-x'|$
at short distances and of a power law behavior, $ - \frac{\tilde{K}_bz_J}{M^2}|x-x'|^{-1}$ at large distances. 
The attractive interaction enhances binding of vortex-antivortex pairs at low temperature,
thus causing enhanced stability of the superconducting phase.
The present study has been made based on the U(1) slave-boson theory concerned with the single holon order parameter. Thus
it will be of great interest to apply our recent SU(2) theory\cite{Lee_Salk} involved with
holon-pair boson order parameter.

We thank Dr. Sung-Sik Lee and Mr. Jae-Gon Eom for helpful discussions.
One of us (SHSS) greatly acknowledges the supports of Korea Ministry of Education (Hakjin program 2003)
and BSRI program (2003) at Pohang University of Science and Technology.

\begin{itemize}
\item[Fig. 1]
The total interaction energy (solid line), logarithmic interaction energy (dashed line) and
additional interaction energy (dotted line) (in the unit of t) at underdoping $\delta \sim 0.035$
as a function of vortex distance d (in the unit of $t^{-1}$)
\end{itemize}

\begin{itemize}
\item[Table 1]
Comparison of the interaction potential between vortices in $s-wave$ and $d-wave$ superconductors
\end{itemize}

\begin{itemize}
\item[Table 2]
Comparison of the interaction potential between vortices in the $d-wave$ BCS formalism and
in the present gauge theory formalism\end{itemize}

\begin{figure}
\vspace{-0cm}
\centerline{\epsfig{file= 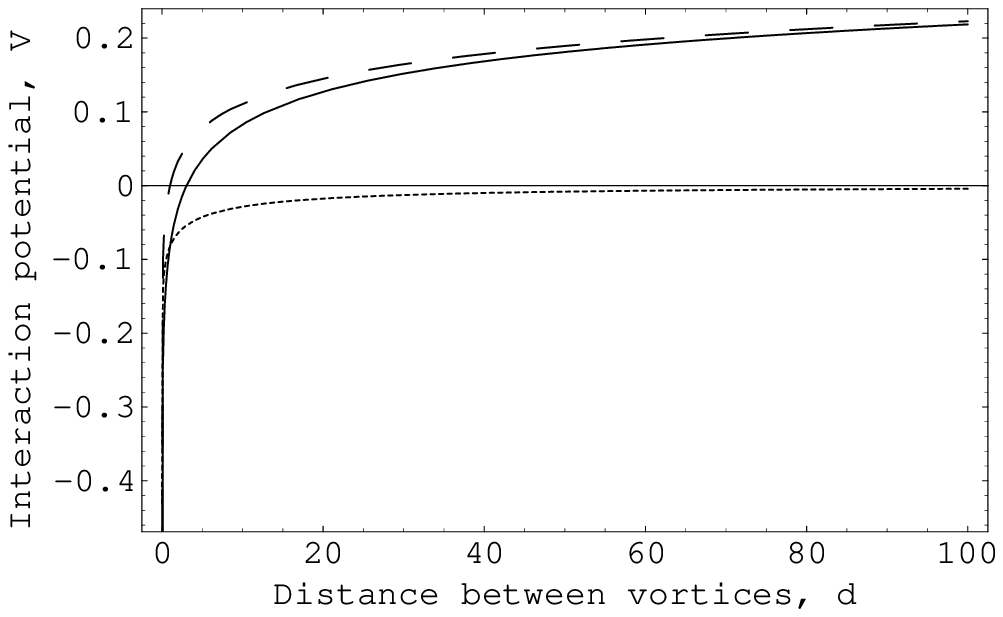,width = 8cm}}
\caption{}
\label{Fig. 1}
\end{figure}

\widetext

\begin{table}
\caption{}
\begin{tabular}{ccccc}
 $s-wave$ superconductors&$d-wave$ superconductors\nn
   $L = \frac{K_p}{2}|\partial\theta_p|^2$ & $L = \frac{K_p}{2}|\partial\theta_p|^2 - iz_JJ_f\partial\theta_p
+ \bar{\psi}_l\gamma\partial\psi_l$ \nn
$L_{dual} = \frac{1}{2K_p}|\partial\times{c}|^2 + icJ_V$ \mbox{  } ($J_{V} \equiv \partial\times\partial\theta_{p}$) &
$L_{dual} = \frac{1}{2K_p}|\partial\times{c}|^2
+ icJ_V + \frac{N}{16}z_J^2J_V\frac{1}{\sqrt{-\partial^2}}J_V$ \nn
$V(q) = \frac{K_p}{q^2}$ & $V(q) = \frac{K_p}{q^2} + \frac{Nz_J^2}{8q}$ \nn
$V(x) = K_pln|x|$&$V(x) = K_pln|x| - \frac{N}{8}z_J^2\frac{1}{|x|}$
\end{tabular}
\end{table}

\begin{table}
\caption{}
\begin{tabular}{ccccc}
 $d-wave$ BCS theory & Our theory\nn
 $L = \frac{K_p}{2}|\partial\theta_p|^2 - iz_JJ_f\partial\theta_p
+ \bar{\psi}_l\gamma\partial\psi_l$ & $L = \frac{\tilde{K}_b}{2}|\partial\theta_p - \tilde{a}|^2
+ \frac{N}{16}(\partial\times\tilde{a})\frac{1}{\sqrt{-\partial^2}}(\partial\times\tilde{a})
+  \frac{1}{2}m_{a}^2\tilde{a}^2$\nn
$L_{dual} = \frac{1}{2K_p}|\partial\times{c}|^2
+ icJ_V + \frac{N}{16}z_J^2J_V\frac{1}{\sqrt{-\partial^2}}J_V$ \mbox{  } ($J_{V} \equiv \partial\times\partial\theta_{p}$) &
 $L_{dual} = \frac{1}{2\tilde{K}_{b}}|\partial\times{c}|^2
+ icJ_V + \frac{1}{2}(\partial\times{c})\frac{1}{\frac{N}{8}\sqrt{-\partial^2} + m_{a}^2}(\partial\times{c})$\nn
$V(q) = \frac{K_p}{q^2} + \frac{Nz_J^2}{8q}$ & $V(q) = \frac{K_p}{q^2}+\tilde{K}_bz_J\frac{1}{q(q+M^2)}$\nn
$V(x) = K_pln|x| - \frac{N}{8}z_J^2\frac{1}{|x|}$& $V(x) = K_pln|x| - \frac{N}{8}z_J^2\frac{1}{|x|}$ \mbox{ for } $q << M^2$ \nn
 & $V(x) = (K_p + \frac{\pi}{2}\tilde{K}_{b}z_{J})ln|x|$ \mbox{ for } $q >> M^2$
\end{tabular}
\end{table}

\narrowtext

\noindent

\widetext


\begin{thebibliography}{9}
\bibitem{Plots} 
The stiffness parameters, $K_b$ and $K_{sp}$, are $K_{b} = \frac{\bar{\rho}_b}{m_b} = 2t\chi_0\delta$ and 
$K_{sp} = J\Delta_0^2 = J(\delta_0 - \delta)^{1.5}$. 
$\chi_0 = J\delta_0^{1.5} + 20\delta$ is the hopping order parameter
(which fits the computed result of the U(1) slave boson mean field theory (Ref. [23])),
where $\delta_0$ satisfies the condition, $T^*(\delta_0) = 0$ ($T^*$ : the PG temperature)
(see the Ref. [9]).
\bibitem{Additional_interaction} After integrating over the dual gauge field $c_{\mu}$ in Eq. [6], we obtain
the total interaction energy between vortices, $\frac{1}{2}(J_{pV} - 2\partial\times{A})_{\mu}P_{\mu\nu}
(J_{pV} - 2\partial\times{A})_{\nu}$, where $J_{pV,\mu} = -i(\psi_{pV}^{\dagger}\partial_{\mu}\psi_{pV} 
- \psi_{pV}\partial_{\mu}\psi_{pV}^{\dagger})$ is the vortex current and $P_{\mu\nu}$, 
the renormalized propagator of the dual gauge field (Eq. [7]). The first term $\frac{K_p}{q^2}$ in Eq. [7] is the same as that
obtained by the first term in Eq. [11] and the second term $\frac{K_b}{4}z_J\frac{1}{q(q+M^2)}$ is equal to the second term in Eq. [11]
in the limit of $q << M^2$ (see the text). Since we neglect the local interaction between Dirac fermions in Eq. [11], the 
second (logarithmic) term in Eq. [8] is not reproduced.
\end{thebibliography}
\end{document}